\documentclass[amsmath,amssymb,amsfonts,aps,prl,twocolumn,superscriptaddress,footinbib]{revtex4-2}

\usepackage{hyperref}
\usepackage{graphicx}
\usepackage[utf8]{inputenc}
\usepackage{amsmath}
\usepackage{comment}
\usepackage{bm}
\usepackage{dsfont}
\usepackage{amsfonts}
\usepackage{xcolor}

\begin{document}

\title{On determining the energy dispersion of spin excitations with scanning tunneling spectroscopy}

\author{J. C. G. Henriques}
\affiliation{International Iberian Nanotechnology Laboratory (INL), Av. Mestre Jos\'e Veiga, 4715-330 Braga, Portugal}
\affiliation{Universidade de Santiago de Compostela, 15782 Santiago de Compostela, Spain}

\author{Chenxiao Zhao}
\affiliation{Empa -- Swiss Federal Laboratories for Materials Science and Technology, D\"{u}bendorf, Switzerland}

\author{G. Catarina}
\affiliation{Empa -- Swiss Federal Laboratories for Materials Science and Technology, D\"{u}bendorf, Switzerland}

\author{Pascal Ruffieux}
\affiliation{Empa -- Swiss Federal Laboratories for Materials Science and Technology, D\"{u}bendorf, Switzerland}

\author{Roman Fasel}
\affiliation{Empa -- Swiss Federal Laboratories for Materials Science and Technology, D\"{u}bendorf, Switzerland}
\affiliation{University of Bern, Bern, Switzerland}

\author{J. Fern\'andez-Rossier}
\altaffiliation[On permanent leave from ]{Departamento de F\'isica Aplicada, Universidad de Alicante, 03690 San Vicente del Raspeig, Spain.}
\affiliation{International Iberian Nanotechnology Laboratory (INL), Av. Mestre Jos\'e Veiga, 4715-330 Braga, Portugal}

\date{\today}


\begin{abstract}
Conventional methods to measure the dispersion relations of collective spin excitations involve probing bulk samples with particles such as neutrons, photons or electrons, which carry a well-defined momentum. Open-ended finite-size spin chains, on the contrary, do not have a well-defined momentum due to the lack of translation symmetry, and their spin excitations are measured with an eminently local probe, using inelastic electron tunneling spectroscopy (IETS) with a scanning tunneling microscope (STM). Here we discuss under what conditions STM-IETS spectra  can be Fourier-transformed to yield dispersion relations in these systems. We relate the success of this approach to the degree to which spin excitations form standing waves. We show that STM-IETS can reveal the energy dispersion of magnons in ferromagnets and triplons in valence bond crystals, but not that of spinons, the spin excitations in Heisenberg spin-1/2 chains. We compare our theoretical predictions with state-of-the-art measurements on nanographene chains that realize the relevant spin Hamiltonians.
\end{abstract}

\maketitle
Understanding the elementary excitations of spin systems is crucial for exploring their dynamics and many-body interactions. In ferromagnetic systems, collective excitations typically take the form of magnons \cite{ pires2021theoretical,Prokop2009, Hisatomi2016, Tabuchi2014, tabuchi2015coherent, chumak2015magnon, Pershoguba2018}, which play a crucial role in spintronics \cite{kubota2008quantitative, sankey2008measurement}. In antiferromagnetic systems, however, a range of distinct excitations may arise. For instance, in a one-dimensional spin-1/2 Heisenberg model, physical excitations form a continuum in the energy-momentum plane due to the excitation of spinon pairs \cite{ pires2021theoretical, kulka2023nature, gauyacq2014excitation, mourigal2013fractional, Muller1981, Lorente2011, Clozieaux1962, hirobe2017one, wu2019tomonaga}. Alternatively, in gapped systems like the alternating-exchange Heisenberg model, triplons, triplet excitations over a valence-bond  crystal state,  have a well defined $E(k)$ dispersion curve \cite{ pires2021theoretical, Hwang2016, Sachdev1990, Kumar2010, Kumar2009, Kumar2008, Gopalan1994}. Both of these excitations emerge over a ground state which lacks long range order, and entangles a macroscopic number of spins \cite{mourigal2013fractional}.

The excitations of artificial spin chains, made on surfaces both with atomic \cite{hirjibehedin2006spin, Choi2019, spinelli2014imaging, wang2024construction,toskovic2016atomic} and molecular \cite{mishra2021observation, zhao2024tunable,sun2024heisenberg, fu2024building, zhao2024gapless, yuan2025fractional, su2024fabrication} spins, can be measured with inelastic electron tunneling spectroscopy (IETS), using  a scanning tunneling microscope (STM). With this method, electrons that tunnel from the tip to the substrate may transfer part of their energy to the spin system, producing magnetic excitations. A promising platform to study quantum magnetism are nanographenes, which, via STM-IETS characterization, have recently been found to realize interesting spin Hamiltonians. Notable examples include the spin-1 Haldane Hamiltonian in [3]-triangulene chains \cite{mishra2021observation}, the spin-1/2 bond-alternating Heisenberg model in chains of Clar’s goblets \cite{zhao2024tunable}, and the spin-1/2 antiferromagnetic Heisenberg model in chains of different molecules \cite{sun2024heisenberg, fu2024building, zhao2024gapless, yuan2025fractional}.

Traditional methods for determining dispersion relations in condensed matter, such as inelastic neutron scattering \cite{Lovesey1984, rossat1991neutron}, Brillouin scattering \cite{Benedek1966, dil1982brillouin} and angle-resolved photoemission spectroscopy \cite{Damascelli2003, Sobota2021}, rely on particles (neutrons, photons, or electrons) with a well defined momentum exciting a target system. This allows for the experimental mapping of an energy vs momentum dispersion relation with these techniques. Contrarily to these approaches, in STM measurements the tunneling electrons do not have a well defined in-plane momentum, which makes it impossible to directly extract a momentum resolved energy dispersion. To circumvent this, spatially resolved STM data have been combined with Fourier analysis to map electronic dispersions in the past \cite{Petersen1998,Pascual2004,Sode2015,Schouteden2009}. Analogously, and motivated by recent experiments in artificial spin chains \cite{zhao2024tunable, su2024fabrication, yuan2025fractional}, we address the question of whether  the site resolved STM-IETS  can be used to infer the dispersion relation of excitations in artificial spin chains. Contrarily to Refs. \cite{Petersen1998,Pascual2004,Sode2015,Schouteden2009}, in this case this possibility is not straightforward, because while in bulk samples the effect of the boundaries is negligible, and therefore the states of the system can be labeled with a wave-vector, for on-surface spin chains the surface to volume ratio is much larger and it is not guaranteed that a momentum labeling of the states is appropriate.

\begin{figure}
    \centering
    \includegraphics[scale = 1]{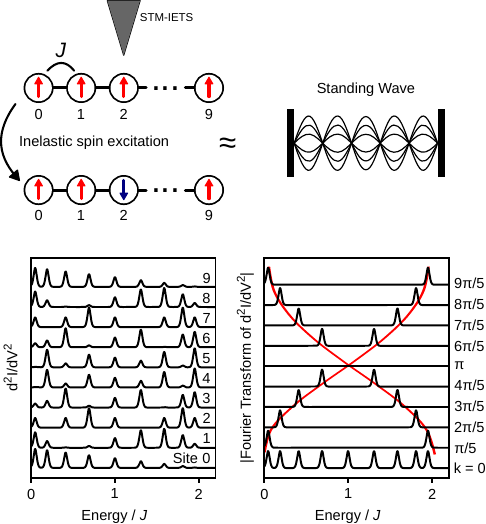}
    \caption{Schematic description of the method proposed to extract the momentum-resolved energy dispersion of spin excitations. Through the use of STM-IETS, it is possible to obtain $d^2I/dV^2$ plots as a function of the applied bias at every site of the spin chain. If the spin excitations probabilities behave (at least approximately) as standing waves, the Fourier transform of the $d^2I/dV^2$ allows us to recover their energy dispersion.
    }
    \label{fig:scheme}
\end{figure}

Here, we establish a theoretical framework to derive momentum-resolved energy dispersion from IETS measurements performed on spin chains with open boundary conditions (OBC), schematically represented in Fig. \ref{fig:scheme}. We show that the feasibility of this approach hinges on how close the chain states are to standing waves formed from the solutions of the associated periodic boundary condition (PBC) system, whose eigenstates are traveling waves with well defined momentum.

One of the key quantities obtained experimentally from IETS is the second derivative of the tunneling current with respect to the applied bias ($d^2I/dV^2$). From second order perturbation theory, one finds the following expression for the site-dependent $d^2I/dV^2$ in the limit where the ground state is unique and the excitation gap is much larger than the thermal energy \cite{Appelbaum1967Exchange, Rossier2009ITS, ternes2015spin}:
\begin{align}
    \frac{d^2I}{dV^2}(l) \propto \sum_M W_M(l) F(eV - E_M + E_\textrm{GS}) \label{eq:d2IdV2}
\end{align}
where $l=0,1,...,L-1$ labels the sites of the spin chain, the sum over $M$ runs over the states of the system, $F(x)$ is a thermally broadened odd delta function, $eV$ is the applied bias energy, and $E_M- E_\textrm{GS}$ is the energy of the $M$ excitation of the system. 
The key quantity in this expression is the spin spectral weight $$W_M(l) = \sum_{a = x,y,z} |\langle \textrm{GS} | S_a (l) | M \rangle|^2,$$ with $S_a(l)$ the $S_a$ spin-operator acting on site $l$, which determines the probability of exciting a given state $|M\rangle$  when the STM tip is placed over the $l$-th site of the chain; this quantity controls the height of the steps in the $dI/dV$ maps across the spin chains, and in that sense provides a real space modulation of the $M$-excitation. 
For simplicity, herein we consider the case of homogeneous spin chains, so that $L$ denotes simultaneously the number of spins and unit cells; for dimerized chains, the index $l$ should run over the unit cells, and an additional index do distinguish between sites within a unit cell must be introduced.
Based on Eq. (\ref{eq:d2IdV2}), the goal of this paper can be phrased as whether the Fourier transform $W_M(k)=\sum_{l=0}^{L-1} e^{-ikl} W_M(l)$ permits us to infer the mapping of the excitations in reciprocal space, and to retrieve an energy dispersion formula $E(k)$.

Through a unitary transformation, we can express the states of a chain (OBC) as linear combination of the eigenstates obtained for the equivalent Hamiltonian in a ring (PBC). As mentioned above, one crucial property of the PBC states is that, due to their translational invariance, we can label them with a given momentum $k$. The value of $k$ is obtained from the diagonalization of the translation operator, defined as $C_L = \prod_{l=0}^{L-2} (1/2 + 2\boldsymbol{S} (l) \cdot \boldsymbol{S} (l+1)$) for spin-1/2 systems,  and can take the values $2\pi n /L$ where $n=0,1,...,L-1$. For the dimerized chain, the values of momenta are obtained from the diagonalization of $C_L^2$ since there are two spins per unit cell. Using this decomposition of OBC states in terms of PBC states to compute $W_M(k)$, we find in general (with details given in \cite{SM}):
\begin{align}
    W_M(k) = \sum_{\lambda_1, \lambda_2, \lambda_3, \lambda_4} &  c_M(k_{\lambda_1})c^*_M(k_{\lambda_2}) c_\textrm{GS}^*(k_{\lambda_3})c_\textrm{GS}(k_{\lambda_4})  \notag \\
    \times & \delta_{k,k_{\lambda_1} - k_{\lambda_3} + k_{\lambda_4} - k_{\lambda_2}} w^{\lambda_3, \lambda_4}_{\lambda_1,\lambda_2}
    \label{eq:general_W_M(k)}
\end{align}
with $w^{\lambda_3,\lambda_4}_{\lambda_1,\lambda_2} = \sum_{a=x,y,z} \langle \lambda_3 | S_a(0) | \lambda_1 \rangle \langle \lambda_2 | S_a(0) | \lambda_4 \rangle$. Here, $|\lambda\rangle$ labels the eigenstates of the Hamiltonian with PBC, which have a well defined momentum $k_\lambda$. The coefficients $c_M(k_\lambda)$ are the overlap between the OBC state $|M\rangle$ and the PBC state $|\lambda\rangle$. The Fourier transform of the second derivative of the current relative to the bias is simply given by $d^2I/dV^2(k) = \sum_M W_M(k) F(eV-E_M+E_{\textrm{GS}})$. Thus, we find that, in general, one can expect contributions to $d^2 I/dV^2(k)$ at every possible momentum $k=k_{\lambda_1} - k_{\lambda_3} + k_{\lambda_4} - k_{\lambda_2}$, with an intensity controlled by the overlap coefficients and the transitions' spectral weight.

To gain insight, let us consider that the OBC and PBC ground states are identical (which we shall see is the case for a ferromagnet), and that any given $|M\rangle$ state is composed of a superposition of only two PBC states, $|\pm\rangle$ with symmetric momenta $\pm k_M$, i.e. the OBC state is a standing wave made of PBC traveling modes. Taking into account these two assumptions and inserting them in Eq. (\ref{eq:general_W_M(k)}) one finds:
\begin{align}
    |W_M(k)| &= \frac{1}{2} \big( \delta_{k,0} | \tilde{w}_{++} + \tilde{w}_{--} | \nonumber \\
     & \quad + \delta_{k,2k_M} |\tilde{w}_{+-}| + \delta_{k,-2k_M} |\tilde{w}_{-+}| \big),
\end{align}
with $\tilde{w}_{\lambda_1,\lambda_2} = \sum_{a=x,y,z} \langle \textrm{GS}|S_a(0) |\lambda_1 \rangle \langle \lambda_2 | S_a(0) | \textrm{GS} \rangle$.
From this expression, we see that when producing a 2D map of energy vs momentum of the $|d^2I / dV^2 (k)|$, we expect peaks at the OBC energies $E_{M}^{\textrm{OBC}}-E_{\textrm{GS}}^{\textrm{OBC}}$ and momenta $k=0,\pm 2k_M$. The excitation energy associated to a transition from the ground state to the standing wave $M$ state in OBC and PBC are related according to $E_{M}^{\textrm{OBC}}-E_{\textrm{GS}}^{\textrm{OBC}} = (1-1/L)(E_{M}^{\textrm{PBC}}-E_{\textrm{GS}}^{\textrm{PBC}})$ (for a dimerized chain, in the strong dimerization limit, the term inversely proportional to the number of unit cells vanishes identically \cite{SM}). Hence, in the thermodynamic limit ($L\gg 1$), $E_{M}^{\textrm{OBC}}-E_{\textrm{GS}}^{\textrm{OBC}} \approx E_{M}^{\textrm{PBC}}-E_{\textrm{GS}}^{\textrm{PBC}}$. Thus, under the considered assumptions, by experimentally obtaining the site resolved differential conductance of OBC spin chains, we can approximately recover the momentum resolved energy dispersion of the magnetic excitations of the system with PBC, with two caveats: 
i) the momentum is doubled for approximately the same energy; 
ii) signal at $k=0$ is always present. 

As we stray away from the simplifying  assumptions, and the standing wave picture is no longer valid, it is not granted that the FT of  $d^2I / dV^2$ can be used to determine the energy dispersion. In order to assess the validity of the method, we consider three kinds of spin-1/2 systems: i) a ferromagnetic spin chain; ii) a dimerized antiferromagnetic spin chain and iii) an antiferromagnetic spin chain, without dimerization. The last two cases will be studied together as we move from one system to the other by controlling the dimerization strength.

\begin{figure}
    \centering
    \includegraphics[width=\linewidth]{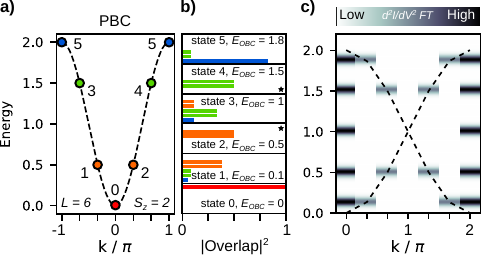}
    \caption{a) Energies (in units of the exchange) of an $L=6$ ferromagnetic Heisenberg spin-1/2 ring, in the $S_z = L/2-1$ sector. Each state is labeled according to the eigenvalue of $C_{L=6}$ (in units of the inverse length of the unit cell). The dashed line is the magnon energy dispersion. b) Absolute squared value of the overlap between the eigenstates of the Hamiltonian with PBC and OBC. States marked with a star symbol are standing waves. c) Color map of $|d^2 I/dV^2(k)|$, obtained for the OBC chain, as a function of momentum and applied bias. The dashed lines are given by $E(k) = 1 - \cos (k/2)$ and a replica shifted by $2\pi$.}
    \label{fig:FM}
\end{figure}

Owning to its simplicity, the ferromagnetic Heisenberg Hamiltonian can be treated fully analytically for a specific $S_z$ sector as we show in \cite{SM}. Here, we will focus on numerical results found for the case with $L=6$ spin-1/2. In the following discussion, all energies are given in units of the ferromagnetic exchange, and the momenta are given in units of the inverse unit cell length. In Fig. \ref{fig:FM}a we display the eigenstates of the PBC Hamiltonian, in the $S_z = L/2 - 1$ sector (corresponding to one magnon excitations), each labeled according to its momentum as found from the diagonalization of the $C_{L=6}$ translation operator. Superimposed with this is the magnon energy dispersion as found from linear spin wave theory, $E(k)=1-\cos k$ \cite{pires2021theoretical}, which perfectly matches the energies found from the diagonalization of the PBC Hamiltonian. 

In Fig. \ref{fig:FM}b we show the decomposition of the 6 eigenstates of the spin Hamiltonian with OBC in terms of the PBC eigenstates, and their respective energies. From this figure, we see that the OBC and PBC ground states are identical (expected for a ferromagnet). As for the excited states, we find that the even-numbered ones are standing waves (marked by a star symbol), while the odd-numbered ones mix more than two PBC states. 

In Fig. \ref{fig:FM}c we show $d^2I/dV^2(k)$ as a contour plot in the $(\textrm{eV}, k)$ plane. As expected, a constant signal at $k=0$ appears at every OBC energy (with a replica shifted by $2\pi$). Furthermore, we find peaks at energies $E = 1/2$ and $E = 3/2$ with momenta $k=2\pi/3$ and $k=4\pi/3$, respectively, as expected from the standing wave character of these states (formed from PBC states with momenta $k=\pi/3$ and $k=2\pi/3$, respectively). For the peaks at $E=0.1$ and $E=1.8$ the only signal appears at $k=\pi/3$ and $k=5\pi/3$. This comes about due to an interference between the different PBC states which form the respective OBC states \cite{SM}; these features, also found in the analytical treatment \cite{SM}, are hard to predict \emph{a priori} as these states are not standing waves. In the center of the Brillouin zone, $k=\pi$, no signal is found; this is a property of the Fourier transform of a signal which is symmetric relatively to the inversion center of the chain \cite{SM}. Overlaid with the $d^2I/dV^2(k)$ color map is the energy dispersion of magnons with $k\rightarrow k/2$ (together with a replica shifted by $2\pi$); we see that the method we propose perfectly traces the theoretical curves for the magnon dispersion. Hence, exploiting the atomic precision of STM-IETS, we can experimentally obtain the momentum resolved energy dispersion of magnons in FM spin chains (this becomes even clearer for larger chains as we show in \cite{SM}). 

\begin{figure}
    \centering
    \includegraphics[scale = 1]{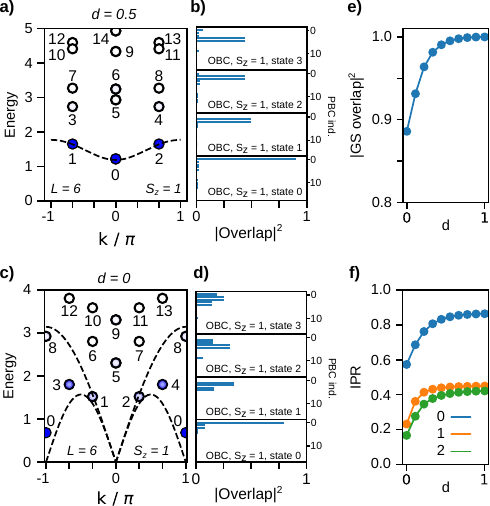}
    \caption{Energies (in units of the exchange) of an $L=6$ spin-1/2 bond alternating antiferromagnetic Heisenberg ring, in the $S_z = 1$ sector, for $d=0.5$ (a) and $d=0$ (c). The momentum of each state is obtained from the eigenvalues of $C_{L=6}$ (in units of the inverse length of the unit cell), and the circle's opacity is proportional to its spin spectral weight. The dashed lines in a) and c) are the triplon \cite{Collins2006} and spinon \cite{Clozieaux1962} energy dispersions, respectively. Absolute square value of the overlap between the eigenstates of the Hamiltonian with PBC and OBC for $d=0.5$ (b) and $d=0$ (d). Absolute square value of the overlap between the OBC and PBC ground states (e) and IPR for the first three OBC excited states (f) as a function of the dimerization.}
    \label{fig:AFM}
\end{figure}

Consider now an antiferromagnetic spin-1/2 Heisenberg model, where we allow for the alternation of exchange between first neighbors. We characterize this alternation through the dimerization parameter $d = (J_1 - J_2) / (J_1 + J_2$), where $J_1 = 1 + d$ and $J_2 = 1-d$ are the exchange values in consecutive bonds; we vary $d$ from 0 to 1. In Figs. \ref{fig:AFM}a and c we show the eigenvalues of the Hamiltonian with PBC and $L=6$, in the $S_z = 1$ sector, with the states labeled according to their momentum $k$, for the cases of $d=0.5$ and $d=0$, respectively. Notice that, because the number of unit cells for $d\neq0$ is half of the one for $d=0$ when the total number of spins is the same, the number of distinct $k$ is also reduced by a factor of 2. 
In Fig. \ref{fig:AFM}a, the dashed line is given by $E(k)=(1+d) - (1-d)/2 \cos k$ and represents the expected energy dispersion of triplons in this system \cite{Collins2006}. In Fig. \ref{fig:AFM}c the dashed lines mark the lower and upper bounds of the spinon continuum, given by $E(k) = \pi/2 |\sin k|$ and $E(k)=\pi |\sin(k/2)|$, respectively \cite{Clozieaux1962}. 

In Figs. \ref{fig:AFM}b and d we depict the decomposition of the first OBC states with $S_z = 1$ in terms of the PBC states in the same $S_z$ sector. These figures reveal that the excited states are closer to the standing wave picture for $d=0.5$ and have a more mixed character when $d=0$, with several PBC states with different momentum contributing to the formation of the OBC states. 

To further understand the role of the dimerization ($d$) in the decomposition of OBC states in terms of their PBC counterparts, we compute the overlap between the OBC and PBC ground states (both being singlets) as well as the inverse participation ratio (IPR) of the OBC excited states, as a function of $d$. We define the IPR for the OBC state $M$ as $\sum_{\lambda} |c_M(k_\lambda)|^4$. Regarding the ground states, we observe that their overlap increases with increasing dimerization. As for the IPR, it approaches 1 with increasing $d$ for the first excited state (implying that we approach a one to one correspondence between OBC and PBC for the first excited state, i.e. a standing wave with $k=0$), and gets closer to 1/2 for the following two excited states (indicating that these states approach a standing wave formed by two traveling waves with symmetric momenta). Based on both of these remarks, we expect that the method we propose works better to map the excitations of a dimerized chain, while its applicability to the homogeneous case is dubious.

\begin{figure*}
    \centering
    \includegraphics[width = \textwidth]{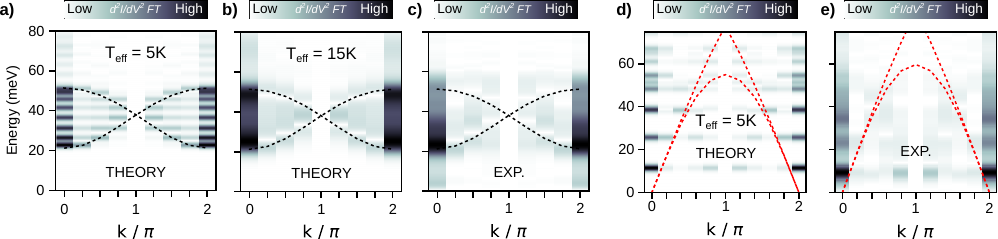}
    \caption{Simulated color maps of $|d^2 I/dV^2(k)|$ for $d=0.24$ for the effective temperatures $T=5$ and $15$K (a and b, respectively) for an open-ended spin-1/2 Heisenberg chain with $L=16$. In panel d) the same is shown for the case without dimerization ($d=0$) for a chain with $L=10$ spins. The dashed lines are obtained from the dispersions given in the text for triplons ($d\neq 0$) and spinons ($d=0$), with $k\rightarrow k/2$, and a replica shifted by $2\pi$, showing good agreement only for the dimerized case (triplons). Panels c and e show the equivalent result obtained experimentally.}
    \label{fig:dispersions}
\end{figure*}

To verify this hypothesis, we perform an exact diagonalization of the alternating exchange Heisenberg antiferromagnetic model for $d=0.24$ and $d=0$ (relevant to compare with experiments \cite{zhao2024tunable, zhao2024gapless,SM}), compute the site dependent $d^2I/dV^2$ and its Fourier transform. In the dimerized case an additional step is needed, where the sum of the two $d^2I/dV^2$ in each unit cell is taken, so as to obtain a unit-cell specific measurement. The results for the two cases are depicted in Fig. \ref{fig:dispersions}. For $d=0.24$, and considering an effective temperature of 5K in the simulation (Fig. \ref{fig:dispersions}a) the triplon energy dispersion is easy to identify, as can be seen by the similar features shared between the $d^2I/dV^2(k)$ color map and the theoretical triplon dispersion line (with $k\rightarrow k/2$). As we increase the effective temperature in the simulation from 5K to 15K (Fig. \ref{fig:dispersions} b), the color map features become broader, but still follows the trend given by the theoretical line. Comparing the simulation with $T_\textrm{eff} = 15$K with the experimental data (panel c) a clear agreement is found. As before, peaks at $0$ and $2\pi$ are visible for all energies, and no signal is found for $k=\pi$. When we remove the dimerization ($d=0$) in  Fig. \ref{fig:dispersions}d, it is no longer true that the features in the color map are well captured by the theoretical bounds for the spinon continuum. Even though some features are shared between the two data sets, other very prominent ones do not match the expected theoretical result. This is also found in the experimental results depicted in Fig. \ref{fig:dispersions}e, which agree well with the simulation shown in panel d, but do not match the spinon theory lines. Thus, as expected from our previous discussion, when applied to the dimerized chain, the method we propose allows us to map the momentum dependent energy dispersion of triplons using STM, but it does not work in the undimerized case.

In summary, we have established the theory to assess under which circumstances the Fourier transform of the $d^2I/dV^2$ can be used to extract the dispersion relation of excitations in spin chains. We found that the crucial condition for the method to work is that the spin excitations behave (at least approximately) as standing waves. For future reference, we name this method as Scanning Tunneling Energy Dispersion Spectroscopy (STEDS). Furthermore, we showed that STEDS works perfectly to study magnons in a 1D ferromagnetic Heisenberg model, and is also suitable to probe triplons in the bond alternating antiferromagnetic Heisenberg spin-1/2 model. At odds with this, STEDS fails to map spinons when the exchange alternation vanishes. Our theory predictions were backed by experimental results obtained from nanographene chains, which have recently been found to realize the desired spin Hamiltonians. Based on the mapping of the ferromagnetic problem to a bosonic one, the proposed method could also be applied to map the energy dispersion of phonons in molecular chains \cite{kittel1991quantum,krane2024vibrational}. 

\section*{Acknowledgments}
This research was supported by the Swiss National Science Foundation (grant No. 212875 and CRSII5 205987), the Werner Siemens Foundation (CarboQuant), the European Union (grant FUNLAYERS-101079184), the Fundação para a Ciência e a Tecnologia (grant number PTDC/FIS-MAC/2045/2021), Generalitat Valenciana (grant number Prometeo2021/017), MICIN-Spain (grant numbers PID2019-109539GB-C41 and PID2022-141712NB-C22) and the Advanced Materials programme supported by MCIN with funding from European Union NextGenerationEU (PRTR-C17.I1)


\begin{thebibliography}{52}%
\makeatletter
\providecommand \@ifxundefined [1]{%
 \@ifx{#1\undefined}
}%
\providecommand \@ifnum [1]{%
 \ifnum #1\expandafter \@firstoftwo
 \else \expandafter \@secondoftwo
 \fi
}%
\providecommand \@ifx [1]{%
 \ifx #1\expandafter \@firstoftwo
 \else \expandafter \@secondoftwo
 \fi
}%
\providecommand \natexlab [1]{#1}%
\providecommand \enquote  [1]{``#1''}%
\providecommand \bibnamefont  [1]{#1}%
\providecommand \bibfnamefont [1]{#1}%
\providecommand \citenamefont [1]{#1}%
\providecommand \href@noop [0]{\@secondoftwo}%
\providecommand \href [0]{\begingroup \@sanitize@url \@href}%
\providecommand \@href[1]{\@@startlink{#1}\@@href}%
\providecommand \@@href[1]{\endgroup#1\@@endlink}%
\providecommand \@sanitize@url [0]{\catcode `\\12\catcode `\$12\catcode
  `\&12\catcode `\#12\catcode `\^12\catcode `\_12\catcode `\%12\relax}%
\providecommand \@@startlink[1]{}%
\providecommand \@@endlink[0]{}%
\providecommand \url  [0]{\begingroup\@sanitize@url \@url }%
\providecommand \@url [1]{\endgroup\@href {#1}{\urlprefix }}%
\providecommand \urlprefix  [0]{URL }%
\providecommand \Eprint [0]{\href }%
\providecommand \doibase [0]{https://doi.org/}%
\providecommand \selectlanguage [0]{\@gobble}%
\providecommand \bibinfo  [0]{\@secondoftwo}%
\providecommand \bibfield  [0]{\@secondoftwo}%
\providecommand \translation [1]{[#1]}%
\providecommand \BibitemOpen [0]{}%
\providecommand \bibitemStop [0]{}%
\providecommand \bibitemNoStop [0]{.\EOS\space}%
\providecommand \EOS [0]{\spacefactor3000\relax}%
\providecommand \BibitemShut  [1]{\csname bibitem#1\endcsname}%
\let\auto@bib@innerbib\@empty
\bibitem [{\citenamefont {Pires}(2021)}]{pires2021theoretical}%
  \BibitemOpen
  \bibfield  {author} {\bibinfo {author} {\bibfnamefont {A.~S.~T.}\
  \bibnamefont {Pires}},\ }\href@noop {} {\emph {\bibinfo {title} {Theoretical
  tools for spin models in magnetic systems}}}\ (\bibinfo  {publisher} {IOP
  Publishing},\ \bibinfo {year} {2021})\BibitemShut {NoStop}%
\bibitem [{\citenamefont {Prokop}\ \emph {et~al.}(2009)\citenamefont {Prokop},
  \citenamefont {Tang}, \citenamefont {Zhang}, \citenamefont {Tudosa},
  \citenamefont {Peixoto}, \citenamefont {Zakeri},\ and\ \citenamefont
  {Kirschner}}]{Prokop2009}%
  \BibitemOpen
  \bibfield  {author} {\bibinfo {author} {\bibfnamefont {J.}~\bibnamefont
  {Prokop}}, \bibinfo {author} {\bibfnamefont {W.~X.}\ \bibnamefont {Tang}},
  \bibinfo {author} {\bibfnamefont {Y.}~\bibnamefont {Zhang}}, \bibinfo
  {author} {\bibfnamefont {I.}~\bibnamefont {Tudosa}}, \bibinfo {author}
  {\bibfnamefont {T.~R.~F.}\ \bibnamefont {Peixoto}}, \bibinfo {author}
  {\bibfnamefont {K.}~\bibnamefont {Zakeri}},\ and\ \bibinfo {author}
  {\bibfnamefont {J.}~\bibnamefont {Kirschner}},\ }\href
  {https://doi.org/10.1103/PhysRevLett.102.177206} {\bibfield  {journal}
  {\bibinfo  {journal} {Phys. Rev. Lett.}\ }\textbf {\bibinfo {volume} {102}},\
  \bibinfo {pages} {177206} (\bibinfo {year} {2009})}\BibitemShut {NoStop}%
\bibitem [{\citenamefont {Hisatomi}\ \emph {et~al.}(2016)\citenamefont
  {Hisatomi}, \citenamefont {Osada}, \citenamefont {Tabuchi}, \citenamefont
  {Ishikawa}, \citenamefont {Noguchi}, \citenamefont {Yamazaki}, \citenamefont
  {Usami},\ and\ \citenamefont {Nakamura}}]{Hisatomi2016}%
  \BibitemOpen
  \bibfield  {author} {\bibinfo {author} {\bibfnamefont {R.}~\bibnamefont
  {Hisatomi}}, \bibinfo {author} {\bibfnamefont {A.}~\bibnamefont {Osada}},
  \bibinfo {author} {\bibfnamefont {Y.}~\bibnamefont {Tabuchi}}, \bibinfo
  {author} {\bibfnamefont {T.}~\bibnamefont {Ishikawa}}, \bibinfo {author}
  {\bibfnamefont {A.}~\bibnamefont {Noguchi}}, \bibinfo {author} {\bibfnamefont
  {R.}~\bibnamefont {Yamazaki}}, \bibinfo {author} {\bibfnamefont
  {K.}~\bibnamefont {Usami}},\ and\ \bibinfo {author} {\bibfnamefont
  {Y.}~\bibnamefont {Nakamura}},\ }\href
  {https://doi.org/10.1103/PhysRevB.93.174427} {\bibfield  {journal} {\bibinfo
  {journal} {Phys. Rev. B}\ }\textbf {\bibinfo {volume} {93}},\ \bibinfo
  {pages} {174427} (\bibinfo {year} {2016})}\BibitemShut {NoStop}%
\bibitem [{\citenamefont {Tabuchi}\ \emph {et~al.}(2014)\citenamefont
  {Tabuchi}, \citenamefont {Ishino}, \citenamefont {Ishikawa}, \citenamefont
  {Yamazaki}, \citenamefont {Usami},\ and\ \citenamefont
  {Nakamura}}]{Tabuchi2014}%
  \BibitemOpen
  \bibfield  {author} {\bibinfo {author} {\bibfnamefont {Y.}~\bibnamefont
  {Tabuchi}}, \bibinfo {author} {\bibfnamefont {S.}~\bibnamefont {Ishino}},
  \bibinfo {author} {\bibfnamefont {T.}~\bibnamefont {Ishikawa}}, \bibinfo
  {author} {\bibfnamefont {R.}~\bibnamefont {Yamazaki}}, \bibinfo {author}
  {\bibfnamefont {K.}~\bibnamefont {Usami}},\ and\ \bibinfo {author}
  {\bibfnamefont {Y.}~\bibnamefont {Nakamura}},\ }\href
  {https://doi.org/10.1103/PhysRevLett.113.083603} {\bibfield  {journal}
  {\bibinfo  {journal} {Phys. Rev. Lett.}\ }\textbf {\bibinfo {volume} {113}},\
  \bibinfo {pages} {083603} (\bibinfo {year} {2014})}\BibitemShut {NoStop}%
\bibitem [{\citenamefont {Tabuchi}\ \emph {et~al.}(2015)\citenamefont
  {Tabuchi}, \citenamefont {Ishino}, \citenamefont {Noguchi}, \citenamefont
  {Ishikawa}, \citenamefont {Yamazaki}, \citenamefont {Usami},\ and\
  \citenamefont {Nakamura}}]{tabuchi2015coherent}%
  \BibitemOpen
  \bibfield  {author} {\bibinfo {author} {\bibfnamefont {Y.}~\bibnamefont
  {Tabuchi}}, \bibinfo {author} {\bibfnamefont {S.}~\bibnamefont {Ishino}},
  \bibinfo {author} {\bibfnamefont {A.}~\bibnamefont {Noguchi}}, \bibinfo
  {author} {\bibfnamefont {T.}~\bibnamefont {Ishikawa}}, \bibinfo {author}
  {\bibfnamefont {R.}~\bibnamefont {Yamazaki}}, \bibinfo {author}
  {\bibfnamefont {K.}~\bibnamefont {Usami}},\ and\ \bibinfo {author}
  {\bibfnamefont {Y.}~\bibnamefont {Nakamura}},\ }\href@noop {} {\bibfield
  {journal} {\bibinfo  {journal} {Science}\ }\textbf {\bibinfo {volume}
  {349}},\ \bibinfo {pages} {405} (\bibinfo {year} {2015})}\BibitemShut
  {NoStop}%
\bibitem [{\citenamefont {Chumak}\ \emph {et~al.}(2015)\citenamefont {Chumak},
  \citenamefont {Vasyuchka}, \citenamefont {Serga},\ and\ \citenamefont
  {Hillebrands}}]{chumak2015magnon}%
  \BibitemOpen
  \bibfield  {author} {\bibinfo {author} {\bibfnamefont {A.~V.}\ \bibnamefont
  {Chumak}}, \bibinfo {author} {\bibfnamefont {V.~I.}\ \bibnamefont
  {Vasyuchka}}, \bibinfo {author} {\bibfnamefont {A.~A.}\ \bibnamefont
  {Serga}},\ and\ \bibinfo {author} {\bibfnamefont {B.}~\bibnamefont
  {Hillebrands}},\ }\href@noop {} {\bibfield  {journal} {\bibinfo  {journal}
  {Nature physics}\ }\textbf {\bibinfo {volume} {11}},\ \bibinfo {pages} {453}
  (\bibinfo {year} {2015})}\BibitemShut {NoStop}%
\bibitem [{\citenamefont {Pershoguba}\ \emph {et~al.}(2018)\citenamefont
  {Pershoguba}, \citenamefont {Banerjee}, \citenamefont {Lashley},
  \citenamefont {Park}, \citenamefont {\AA{}gren}, \citenamefont {Aeppli},\
  and\ \citenamefont {Balatsky}}]{Pershoguba2018}%
  \BibitemOpen
  \bibfield  {author} {\bibinfo {author} {\bibfnamefont {S.~S.}\ \bibnamefont
  {Pershoguba}}, \bibinfo {author} {\bibfnamefont {S.}~\bibnamefont
  {Banerjee}}, \bibinfo {author} {\bibfnamefont {J.~C.}\ \bibnamefont
  {Lashley}}, \bibinfo {author} {\bibfnamefont {J.}~\bibnamefont {Park}},
  \bibinfo {author} {\bibfnamefont {H.}~\bibnamefont {\AA{}gren}}, \bibinfo
  {author} {\bibfnamefont {G.}~\bibnamefont {Aeppli}},\ and\ \bibinfo {author}
  {\bibfnamefont {A.~V.}\ \bibnamefont {Balatsky}},\ }\href
  {https://doi.org/10.1103/PhysRevX.8.011010} {\bibfield  {journal} {\bibinfo
  {journal} {Phys. Rev. X}\ }\textbf {\bibinfo {volume} {8}},\ \bibinfo {pages}
  {011010} (\bibinfo {year} {2018})}\BibitemShut {NoStop}%
\bibitem [{\citenamefont {Kubota}\ \emph {et~al.}(2008)\citenamefont {Kubota},
  \citenamefont {Fukushima}, \citenamefont {Yakushiji}, \citenamefont
  {Nagahama}, \citenamefont {Yuasa}, \citenamefont {Ando}, \citenamefont
  {Maehara}, \citenamefont {Nagamine}, \citenamefont {Tsunekawa}, \citenamefont
  {Djayaprawira}, \citenamefont {Watanabe},\ and\ \citenamefont
  {Suzuki}}]{kubota2008quantitative}%
  \BibitemOpen
  \bibfield  {author} {\bibinfo {author} {\bibfnamefont {H.}~\bibnamefont
  {Kubota}}, \bibinfo {author} {\bibfnamefont {A.}~\bibnamefont {Fukushima}},
  \bibinfo {author} {\bibfnamefont {K.}~\bibnamefont {Yakushiji}}, \bibinfo
  {author} {\bibfnamefont {T.}~\bibnamefont {Nagahama}}, \bibinfo {author}
  {\bibfnamefont {S.}~\bibnamefont {Yuasa}}, \bibinfo {author} {\bibfnamefont
  {K.}~\bibnamefont {Ando}}, \bibinfo {author} {\bibfnamefont {H.}~\bibnamefont
  {Maehara}}, \bibinfo {author} {\bibfnamefont {Y.}~\bibnamefont {Nagamine}},
  \bibinfo {author} {\bibfnamefont {K.}~\bibnamefont {Tsunekawa}}, \bibinfo
  {author} {\bibfnamefont {D.~D.}\ \bibnamefont {Djayaprawira}}, \bibinfo
  {author} {\bibfnamefont {N.}~\bibnamefont {Watanabe}},\ and\ \bibinfo
  {author} {\bibfnamefont {Y.}~\bibnamefont {Suzuki}},\ }\href@noop {}
  {\bibfield  {journal} {\bibinfo  {journal} {Nature Physics}\ }\textbf
  {\bibinfo {volume} {4}},\ \bibinfo {pages} {37} (\bibinfo {year}
  {2008})}\BibitemShut {NoStop}%
\bibitem [{\citenamefont {Sankey}\ \emph {et~al.}(2008)\citenamefont {Sankey},
  \citenamefont {Cui}, \citenamefont {Sun}, \citenamefont {Slonczewski},
  \citenamefont {Buhrman},\ and\ \citenamefont
  {Ralph}}]{sankey2008measurement}%
  \BibitemOpen
  \bibfield  {author} {\bibinfo {author} {\bibfnamefont {J.~C.}\ \bibnamefont
  {Sankey}}, \bibinfo {author} {\bibfnamefont {Y.-T.}\ \bibnamefont {Cui}},
  \bibinfo {author} {\bibfnamefont {J.~Z.}\ \bibnamefont {Sun}}, \bibinfo
  {author} {\bibfnamefont {J.~C.}\ \bibnamefont {Slonczewski}}, \bibinfo
  {author} {\bibfnamefont {R.~A.}\ \bibnamefont {Buhrman}},\ and\ \bibinfo
  {author} {\bibfnamefont {D.~C.}\ \bibnamefont {Ralph}},\ }\href@noop {}
  {\bibfield  {journal} {\bibinfo  {journal} {Nature Physics}\ }\textbf
  {\bibinfo {volume} {4}},\ \bibinfo {pages} {67} (\bibinfo {year}
  {2008})}\BibitemShut {NoStop}%
\bibitem [{\citenamefont {Kulka}\ \emph {et~al.}(2023)\citenamefont {Kulka},
  \citenamefont {Panfil}, \citenamefont {Berciu},\ and\ \citenamefont
  {Wohlfeld}}]{kulka2023nature}%
  \BibitemOpen
  \bibfield  {author} {\bibinfo {author} {\bibfnamefont {T.}~\bibnamefont
  {Kulka}}, \bibinfo {author} {\bibfnamefont {M.}~\bibnamefont {Panfil}},
  \bibinfo {author} {\bibfnamefont {M.}~\bibnamefont {Berciu}},\ and\ \bibinfo
  {author} {\bibfnamefont {K.}~\bibnamefont {Wohlfeld}},\ }\href@noop {}
  {\bibfield  {journal} {\bibinfo  {journal} {arXiv preprint arXiv:2303.02276}\
  } (\bibinfo {year} {2023})}\BibitemShut {NoStop}%
\bibitem [{\citenamefont {Gauyacq}\ and\ \citenamefont
  {Lorente}(2014)}]{gauyacq2014excitation}%
  \BibitemOpen
  \bibfield  {author} {\bibinfo {author} {\bibfnamefont {J.}~\bibnamefont
  {Gauyacq}}\ and\ \bibinfo {author} {\bibfnamefont {N.}~\bibnamefont
  {Lorente}},\ }\href@noop {} {\bibfield  {journal} {\bibinfo  {journal}
  {Journal of Physics: Condensed Matter}\ }\textbf {\bibinfo {volume} {26}},\
  \bibinfo {pages} {394005} (\bibinfo {year} {2014})}\BibitemShut {NoStop}%
\bibitem [{\citenamefont {Mourigal}\ \emph {et~al.}(2013)\citenamefont
  {Mourigal}, \citenamefont {Enderle}, \citenamefont {Kl{\"o}pperpieper},
  \citenamefont {Caux}, \citenamefont {Stunault},\ and\ \citenamefont
  {R{\o}nnow}}]{mourigal2013fractional}%
  \BibitemOpen
  \bibfield  {author} {\bibinfo {author} {\bibfnamefont {M.}~\bibnamefont
  {Mourigal}}, \bibinfo {author} {\bibfnamefont {M.}~\bibnamefont {Enderle}},
  \bibinfo {author} {\bibfnamefont {A.}~\bibnamefont {Kl{\"o}pperpieper}},
  \bibinfo {author} {\bibfnamefont {J.-S.}\ \bibnamefont {Caux}}, \bibinfo
  {author} {\bibfnamefont {A.}~\bibnamefont {Stunault}},\ and\ \bibinfo
  {author} {\bibfnamefont {H.~M.}\ \bibnamefont {R{\o}nnow}},\ }\href@noop {}
  {\bibfield  {journal} {\bibinfo  {journal} {Nature Physics}\ }\textbf
  {\bibinfo {volume} {9}},\ \bibinfo {pages} {435} (\bibinfo {year}
  {2013})}\BibitemShut {NoStop}%
\bibitem [{\citenamefont {M\"uller}\ \emph {et~al.}(1981)\citenamefont
  {M\"uller}, \citenamefont {Thomas}, \citenamefont {Beck},\ and\ \citenamefont
  {Bonner}}]{Muller1981}%
  \BibitemOpen
  \bibfield  {author} {\bibinfo {author} {\bibfnamefont {G.}~\bibnamefont
  {M\"uller}}, \bibinfo {author} {\bibfnamefont {H.}~\bibnamefont {Thomas}},
  \bibinfo {author} {\bibfnamefont {H.}~\bibnamefont {Beck}},\ and\ \bibinfo
  {author} {\bibfnamefont {J.~C.}\ \bibnamefont {Bonner}},\ }\href
  {https://doi.org/10.1103/PhysRevB.24.1429} {\bibfield  {journal} {\bibinfo
  {journal} {Phys. Rev. B}\ }\textbf {\bibinfo {volume} {24}},\ \bibinfo
  {pages} {1429} (\bibinfo {year} {1981})}\BibitemShut {NoStop}%
\bibitem [{\citenamefont {Gauyacq}\ and\ \citenamefont
  {Lorente}(2011)}]{Lorente2011}%
  \BibitemOpen
  \bibfield  {author} {\bibinfo {author} {\bibfnamefont {J.~P.}\ \bibnamefont
  {Gauyacq}}\ and\ \bibinfo {author} {\bibfnamefont {N.}~\bibnamefont
  {Lorente}},\ }\href {https://doi.org/10.1103/PhysRevB.83.035418} {\bibfield
  {journal} {\bibinfo  {journal} {Phys. Rev. B}\ }\textbf {\bibinfo {volume}
  {83}},\ \bibinfo {pages} {035418} (\bibinfo {year} {2011})}\BibitemShut
  {NoStop}%
\bibitem [{\citenamefont {des Cloizeaux}\ and\ \citenamefont
  {Pearson}(1962)}]{Clozieaux1962}%
  \BibitemOpen
  \bibfield  {author} {\bibinfo {author} {\bibfnamefont {J.}~\bibnamefont {des
  Cloizeaux}}\ and\ \bibinfo {author} {\bibfnamefont {J.~J.}\ \bibnamefont
  {Pearson}},\ }\href {https://doi.org/10.1103/PhysRev.128.2131} {\bibfield
  {journal} {\bibinfo  {journal} {Phys. Rev.}\ }\textbf {\bibinfo {volume}
  {128}},\ \bibinfo {pages} {2131} (\bibinfo {year} {1962})}\BibitemShut
  {NoStop}%
\bibitem [{\citenamefont {Hirobe}\ \emph {et~al.}(2017)\citenamefont {Hirobe},
  \citenamefont {Sato}, \citenamefont {Kawamata}, \citenamefont {Shiomi},
  \citenamefont {Uchida}, \citenamefont {Iguchi}, \citenamefont {Koike},
  \citenamefont {Maekawa},\ and\ \citenamefont {Saitoh}}]{hirobe2017one}%
  \BibitemOpen
  \bibfield  {author} {\bibinfo {author} {\bibfnamefont {D.}~\bibnamefont
  {Hirobe}}, \bibinfo {author} {\bibfnamefont {M.}~\bibnamefont {Sato}},
  \bibinfo {author} {\bibfnamefont {T.}~\bibnamefont {Kawamata}}, \bibinfo
  {author} {\bibfnamefont {Y.}~\bibnamefont {Shiomi}}, \bibinfo {author}
  {\bibfnamefont {K.-i.}\ \bibnamefont {Uchida}}, \bibinfo {author}
  {\bibfnamefont {R.}~\bibnamefont {Iguchi}}, \bibinfo {author} {\bibfnamefont
  {Y.}~\bibnamefont {Koike}}, \bibinfo {author} {\bibfnamefont
  {S.}~\bibnamefont {Maekawa}},\ and\ \bibinfo {author} {\bibfnamefont
  {E.}~\bibnamefont {Saitoh}},\ }\href@noop {} {\bibfield  {journal} {\bibinfo
  {journal} {Nature Physics}\ }\textbf {\bibinfo {volume} {13}},\ \bibinfo
  {pages} {30} (\bibinfo {year} {2017})}\BibitemShut {NoStop}%
\bibitem [{\citenamefont {Wu}\ \emph {et~al.}(2019)\citenamefont {Wu},
  \citenamefont {Nikitin}, \citenamefont {Wang}, \citenamefont {Zhu},
  \citenamefont {Batista}, \citenamefont {Tsvelik}, \citenamefont {Samarakoon},
  \citenamefont {Tennant}, \citenamefont {Brando}, \citenamefont {Vasylechko},
  \citenamefont {Frontzek}, \citenamefont {Savici}, \citenamefont {Sala},
  \citenamefont {Ehlers}, \citenamefont {Christianson}, \citenamefont
  {Lumsden},\ and\ \citenamefont {Podlesnyak}}]{wu2019tomonaga}%
  \BibitemOpen
  \bibfield  {author} {\bibinfo {author} {\bibfnamefont {L.}~\bibnamefont
  {Wu}}, \bibinfo {author} {\bibfnamefont {S.}~\bibnamefont {Nikitin}},
  \bibinfo {author} {\bibfnamefont {Z.}~\bibnamefont {Wang}}, \bibinfo {author}
  {\bibfnamefont {W.}~\bibnamefont {Zhu}}, \bibinfo {author} {\bibfnamefont
  {C.~D.}\ \bibnamefont {Batista}}, \bibinfo {author} {\bibfnamefont
  {A.}~\bibnamefont {Tsvelik}}, \bibinfo {author} {\bibfnamefont {A.~M.}\
  \bibnamefont {Samarakoon}}, \bibinfo {author} {\bibfnamefont {D.~A.}\
  \bibnamefont {Tennant}}, \bibinfo {author} {\bibfnamefont {M.}~\bibnamefont
  {Brando}}, \bibinfo {author} {\bibfnamefont {L.}~\bibnamefont {Vasylechko}},
  \bibinfo {author} {\bibfnamefont {M.}~\bibnamefont {Frontzek}}, \bibinfo
  {author} {\bibfnamefont {A.~T.}\ \bibnamefont {Savici}}, \bibinfo {author}
  {\bibfnamefont {G.}~\bibnamefont {Sala}}, \bibinfo {author} {\bibfnamefont
  {G.}~\bibnamefont {Ehlers}}, \bibinfo {author} {\bibfnamefont {A.~D.}\
  \bibnamefont {Christianson}}, \bibinfo {author} {\bibfnamefont {M.~D.}\
  \bibnamefont {Lumsden}},\ and\ \bibinfo {author} {\bibfnamefont
  {A.}~\bibnamefont {Podlesnyak}},\ }\href@noop {} {\bibfield  {journal}
  {\bibinfo  {journal} {Nature communications}\ }\textbf {\bibinfo {volume}
  {10}},\ \bibinfo {pages} {698} (\bibinfo {year} {2019})}\BibitemShut
  {NoStop}%
\bibitem [{\citenamefont {Hwang}\ and\ \citenamefont {Kim}(2016)}]{Hwang2016}%
  \BibitemOpen
  \bibfield  {author} {\bibinfo {author} {\bibfnamefont {K.}~\bibnamefont
  {Hwang}}\ and\ \bibinfo {author} {\bibfnamefont {Y.~B.}\ \bibnamefont
  {Kim}},\ }\href {https://doi.org/10.1103/PhysRevB.93.235130} {\bibfield
  {journal} {\bibinfo  {journal} {Phys. Rev. B}\ }\textbf {\bibinfo {volume}
  {93}},\ \bibinfo {pages} {235130} (\bibinfo {year} {2016})}\BibitemShut
  {NoStop}%
\bibitem [{\citenamefont {Sachdev}\ and\ \citenamefont
  {Bhatt}(1990)}]{Sachdev1990}%
  \BibitemOpen
  \bibfield  {author} {\bibinfo {author} {\bibfnamefont {S.}~\bibnamefont
  {Sachdev}}\ and\ \bibinfo {author} {\bibfnamefont {R.~N.}\ \bibnamefont
  {Bhatt}},\ }\href {https://doi.org/10.1103/PhysRevB.41.9323} {\bibfield
  {journal} {\bibinfo  {journal} {Phys. Rev. B}\ }\textbf {\bibinfo {volume}
  {41}},\ \bibinfo {pages} {9323} (\bibinfo {year} {1990})}\BibitemShut
  {NoStop}%
\bibitem [{\citenamefont {Kumar}(2010)}]{Kumar2010}%
  \BibitemOpen
  \bibfield  {author} {\bibinfo {author} {\bibfnamefont {B.}~\bibnamefont
  {Kumar}},\ }\href {https://doi.org/10.1103/PhysRevB.82.054404} {\bibfield
  {journal} {\bibinfo  {journal} {Phys. Rev. B}\ }\textbf {\bibinfo {volume}
  {82}},\ \bibinfo {pages} {054404} (\bibinfo {year} {2010})}\BibitemShut
  {NoStop}%
\bibitem [{\citenamefont {Kumar}\ \emph {et~al.}(2009)\citenamefont {Kumar},
  \citenamefont {Kumar},\ and\ \citenamefont {Kumar}}]{Kumar2009}%
  \BibitemOpen
  \bibfield  {author} {\bibinfo {author} {\bibfnamefont {R.}~\bibnamefont
  {Kumar}}, \bibinfo {author} {\bibfnamefont {D.}~\bibnamefont {Kumar}},\ and\
  \bibinfo {author} {\bibfnamefont {B.}~\bibnamefont {Kumar}},\ }\href
  {https://doi.org/10.1103/PhysRevB.80.214428} {\bibfield  {journal} {\bibinfo
  {journal} {Phys. Rev. B}\ }\textbf {\bibinfo {volume} {80}},\ \bibinfo
  {pages} {214428} (\bibinfo {year} {2009})}\BibitemShut {NoStop}%
\bibitem [{\citenamefont {Kumar}\ and\ \citenamefont
  {Kumar}(2008)}]{Kumar2008}%
  \BibitemOpen
  \bibfield  {author} {\bibinfo {author} {\bibfnamefont {R.}~\bibnamefont
  {Kumar}}\ and\ \bibinfo {author} {\bibfnamefont {B.}~\bibnamefont {Kumar}},\
  }\href {https://doi.org/10.1103/PhysRevB.77.144413} {\bibfield  {journal}
  {\bibinfo  {journal} {Phys. Rev. B}\ }\textbf {\bibinfo {volume} {77}},\
  \bibinfo {pages} {144413} (\bibinfo {year} {2008})}\BibitemShut {NoStop}%
\bibitem [{\citenamefont {Gopalan}\ \emph {et~al.}(1994)\citenamefont
  {Gopalan}, \citenamefont {Rice},\ and\ \citenamefont
  {Sigrist}}]{Gopalan1994}%
  \BibitemOpen
  \bibfield  {author} {\bibinfo {author} {\bibfnamefont {S.}~\bibnamefont
  {Gopalan}}, \bibinfo {author} {\bibfnamefont {T.~M.}\ \bibnamefont {Rice}},\
  and\ \bibinfo {author} {\bibfnamefont {M.}~\bibnamefont {Sigrist}},\ }\href
  {https://doi.org/10.1103/PhysRevB.49.8901} {\bibfield  {journal} {\bibinfo
  {journal} {Phys. Rev. B}\ }\textbf {\bibinfo {volume} {49}},\ \bibinfo
  {pages} {8901} (\bibinfo {year} {1994})}\BibitemShut {NoStop}%
\bibitem [{\citenamefont {Hirjibehedin}\ \emph {et~al.}(2006)\citenamefont
  {Hirjibehedin}, \citenamefont {Lutz},\ and\ \citenamefont
  {Heinrich}}]{hirjibehedin2006spin}%
  \BibitemOpen
  \bibfield  {author} {\bibinfo {author} {\bibfnamefont {C.~F.}\ \bibnamefont
  {Hirjibehedin}}, \bibinfo {author} {\bibfnamefont {C.~P.}\ \bibnamefont
  {Lutz}},\ and\ \bibinfo {author} {\bibfnamefont {A.~J.}\ \bibnamefont
  {Heinrich}},\ }\href@noop {} {\bibfield  {journal} {\bibinfo  {journal}
  {Science}\ }\textbf {\bibinfo {volume} {312}},\ \bibinfo {pages} {1021}
  (\bibinfo {year} {2006})}\BibitemShut {NoStop}%
\bibitem [{\citenamefont {Choi}\ \emph {et~al.}(2019)\citenamefont {Choi},
  \citenamefont {Lorente}, \citenamefont {Wiebe}, \citenamefont {von Bergmann},
  \citenamefont {Otte},\ and\ \citenamefont {Heinrich}}]{Choi2019}%
  \BibitemOpen
  \bibfield  {author} {\bibinfo {author} {\bibfnamefont {D.-J.}\ \bibnamefont
  {Choi}}, \bibinfo {author} {\bibfnamefont {N.}~\bibnamefont {Lorente}},
  \bibinfo {author} {\bibfnamefont {J.}~\bibnamefont {Wiebe}}, \bibinfo
  {author} {\bibfnamefont {K.}~\bibnamefont {von Bergmann}}, \bibinfo {author}
  {\bibfnamefont {A.~F.}\ \bibnamefont {Otte}},\ and\ \bibinfo {author}
  {\bibfnamefont {A.~J.}\ \bibnamefont {Heinrich}},\ }\href
  {https://doi.org/10.1103/RevModPhys.91.041001} {\bibfield  {journal}
  {\bibinfo  {journal} {Rev. Mod. Phys.}\ }\textbf {\bibinfo {volume} {91}},\
  \bibinfo {pages} {041001} (\bibinfo {year} {2019})}\BibitemShut {NoStop}%
\bibitem [{\citenamefont {Spinelli}\ \emph {et~al.}(2014)\citenamefont
  {Spinelli}, \citenamefont {Bryant}, \citenamefont {Delgado}, \citenamefont
  {Fern{\'a}ndez-Rossier},\ and\ \citenamefont {Otte}}]{spinelli2014imaging}%
  \BibitemOpen
  \bibfield  {author} {\bibinfo {author} {\bibfnamefont {A.}~\bibnamefont
  {Spinelli}}, \bibinfo {author} {\bibfnamefont {B.}~\bibnamefont {Bryant}},
  \bibinfo {author} {\bibfnamefont {F.}~\bibnamefont {Delgado}}, \bibinfo
  {author} {\bibfnamefont {J.}~\bibnamefont {Fern{\'a}ndez-Rossier}},\ and\
  \bibinfo {author} {\bibfnamefont {A.~F.}\ \bibnamefont {Otte}},\ }\href@noop
  {} {\bibfield  {journal} {\bibinfo  {journal} {Nature materials}\ }\textbf
  {\bibinfo {volume} {13}},\ \bibinfo {pages} {782} (\bibinfo {year}
  {2014})}\BibitemShut {NoStop}%
\bibitem [{\citenamefont {Wang}\ \emph {et~al.}(2024)\citenamefont {Wang},
  \citenamefont {Fan}, \citenamefont {Chen}, \citenamefont {Jiang},
  \citenamefont {Gao}, \citenamefont {Lado},\ and\ \citenamefont
  {Yang}}]{wang2024construction}%
  \BibitemOpen
  \bibfield  {author} {\bibinfo {author} {\bibfnamefont {H.}~\bibnamefont
  {Wang}}, \bibinfo {author} {\bibfnamefont {P.}~\bibnamefont {Fan}}, \bibinfo
  {author} {\bibfnamefont {J.}~\bibnamefont {Chen}}, \bibinfo {author}
  {\bibfnamefont {L.}~\bibnamefont {Jiang}}, \bibinfo {author} {\bibfnamefont
  {H.-J.}\ \bibnamefont {Gao}}, \bibinfo {author} {\bibfnamefont {J.~L.}\
  \bibnamefont {Lado}},\ and\ \bibinfo {author} {\bibfnamefont
  {K.}~\bibnamefont {Yang}},\ }\href@noop {} {\bibfield  {journal} {\bibinfo
  {journal} {Nature Nanotechnology}\ ,\ \bibinfo {pages} {1}} (\bibinfo {year}
  {2024})}\BibitemShut {NoStop}%
\bibitem [{\citenamefont {Toskovic}\ \emph {et~al.}(2016)\citenamefont
  {Toskovic}, \citenamefont {Van Den~Berg}, \citenamefont {Spinelli},
  \citenamefont {Eliens}, \citenamefont {Van Den~Toorn}, \citenamefont
  {Bryant}, \citenamefont {Caux},\ and\ \citenamefont
  {Otte}}]{toskovic2016atomic}%
  \BibitemOpen
  \bibfield  {author} {\bibinfo {author} {\bibfnamefont {R.}~\bibnamefont
  {Toskovic}}, \bibinfo {author} {\bibfnamefont {R.}~\bibnamefont {Van
  Den~Berg}}, \bibinfo {author} {\bibfnamefont {A.}~\bibnamefont {Spinelli}},
  \bibinfo {author} {\bibfnamefont {I.}~\bibnamefont {Eliens}}, \bibinfo
  {author} {\bibfnamefont {B.}~\bibnamefont {Van Den~Toorn}}, \bibinfo {author}
  {\bibfnamefont {B.}~\bibnamefont {Bryant}}, \bibinfo {author} {\bibfnamefont
  {J.-S.}\ \bibnamefont {Caux}},\ and\ \bibinfo {author} {\bibfnamefont
  {A.}~\bibnamefont {Otte}},\ }\href@noop {} {\bibfield  {journal} {\bibinfo
  {journal} {Nature Physics}\ }\textbf {\bibinfo {volume} {12}},\ \bibinfo
  {pages} {656} (\bibinfo {year} {2016})}\BibitemShut {NoStop}%
\bibitem [{\citenamefont {Mishra}\ \emph {et~al.}(2021)\citenamefont {Mishra},
  \citenamefont {Catarina}, \citenamefont {Wu}, \citenamefont {Ortiz},
  \citenamefont {Jacob}, \citenamefont {Eimre}, \citenamefont {Ma},
  \citenamefont {Pignedoli}, \citenamefont {Feng}, \citenamefont {Ruffieux}
  \emph {et~al.}}]{mishra2021observation}%
  \BibitemOpen
  \bibfield  {author} {\bibinfo {author} {\bibfnamefont {S.}~\bibnamefont
  {Mishra}}, \bibinfo {author} {\bibfnamefont {G.}~\bibnamefont {Catarina}},
  \bibinfo {author} {\bibfnamefont {F.}~\bibnamefont {Wu}}, \bibinfo {author}
  {\bibfnamefont {R.}~\bibnamefont {Ortiz}}, \bibinfo {author} {\bibfnamefont
  {D.}~\bibnamefont {Jacob}}, \bibinfo {author} {\bibfnamefont
  {K.}~\bibnamefont {Eimre}}, \bibinfo {author} {\bibfnamefont
  {J.}~\bibnamefont {Ma}}, \bibinfo {author} {\bibfnamefont {C.~A.}\
  \bibnamefont {Pignedoli}}, \bibinfo {author} {\bibfnamefont {X.}~\bibnamefont
  {Feng}}, \bibinfo {author} {\bibfnamefont {P.}~\bibnamefont {Ruffieux}},
  \emph {et~al.},\ }\href@noop {} {\bibfield  {journal} {\bibinfo  {journal}
  {Nature}\ }\textbf {\bibinfo {volume} {598}},\ \bibinfo {pages} {287}
  (\bibinfo {year} {2021})}\BibitemShut {NoStop}%
\bibitem [{\citenamefont {Zhao}\ \emph
  {et~al.}(2024{\natexlab{a}})\citenamefont {Zhao}, \citenamefont {Catarina},
  \citenamefont {Zhang}, \citenamefont {Henriques}, \citenamefont {Yang},
  \citenamefont {Ma}, \citenamefont {Feng}, \citenamefont {Gr{\"o}ning},
  \citenamefont {Ruffieux}, \citenamefont {Fern{\'a}ndez-Rossier},\ and\
  \citenamefont {Fasel}}]{zhao2024tunable}%
  \BibitemOpen
  \bibfield  {author} {\bibinfo {author} {\bibfnamefont {C.}~\bibnamefont
  {Zhao}}, \bibinfo {author} {\bibfnamefont {G.}~\bibnamefont {Catarina}},
  \bibinfo {author} {\bibfnamefont {J.-J.}\ \bibnamefont {Zhang}}, \bibinfo
  {author} {\bibfnamefont {J.~C.}\ \bibnamefont {Henriques}}, \bibinfo {author}
  {\bibfnamefont {L.}~\bibnamefont {Yang}}, \bibinfo {author} {\bibfnamefont
  {J.}~\bibnamefont {Ma}}, \bibinfo {author} {\bibfnamefont {X.}~\bibnamefont
  {Feng}}, \bibinfo {author} {\bibfnamefont {O.}~\bibnamefont {Gr{\"o}ning}},
  \bibinfo {author} {\bibfnamefont {P.}~\bibnamefont {Ruffieux}}, \bibinfo
  {author} {\bibfnamefont {J.}~\bibnamefont {Fern{\'a}ndez-Rossier}},\ and\
  \bibinfo {author} {\bibfnamefont {R.}~\bibnamefont {Fasel}},\ }\href@noop {}
  {\bibfield  {journal} {\bibinfo  {journal} {Nature Nanotechnology}\ ,\
  \bibinfo {pages} {1}} (\bibinfo {year} {2024}{\natexlab{a}})}\BibitemShut
  {NoStop}%
\bibitem [{\citenamefont {Sun}\ \emph {et~al.}(2024)\citenamefont {Sun},
  \citenamefont {Cao}, \citenamefont {Silveira}, \citenamefont {Fumega},
  \citenamefont {Hanindita}, \citenamefont {Ito}, \citenamefont {Lado},
  \citenamefont {Liljeroth}, \citenamefont {Foster},\ and\ \citenamefont
  {Kawai}}]{sun2024heisenberg}%
  \BibitemOpen
  \bibfield  {author} {\bibinfo {author} {\bibfnamefont {K.}~\bibnamefont
  {Sun}}, \bibinfo {author} {\bibfnamefont {N.}~\bibnamefont {Cao}}, \bibinfo
  {author} {\bibfnamefont {O.~J.}\ \bibnamefont {Silveira}}, \bibinfo {author}
  {\bibfnamefont {A.~O.}\ \bibnamefont {Fumega}}, \bibinfo {author}
  {\bibfnamefont {F.}~\bibnamefont {Hanindita}}, \bibinfo {author}
  {\bibfnamefont {S.}~\bibnamefont {Ito}}, \bibinfo {author} {\bibfnamefont
  {J.~L.}\ \bibnamefont {Lado}}, \bibinfo {author} {\bibfnamefont
  {P.}~\bibnamefont {Liljeroth}}, \bibinfo {author} {\bibfnamefont {A.~S.}\
  \bibnamefont {Foster}},\ and\ \bibinfo {author} {\bibfnamefont
  {S.}~\bibnamefont {Kawai}},\ }\href@noop {} {\bibfield  {journal} {\bibinfo
  {journal} {arXiv preprint arXiv:2407.02142}\ } (\bibinfo {year}
  {2024})}\BibitemShut {NoStop}%
\bibitem [{\citenamefont {Fu}\ \emph {et~al.}(2024)\citenamefont {Fu},
  \citenamefont {Huang}, \citenamefont {Liu}, \citenamefont {Henriques},
  \citenamefont {Gao}, \citenamefont {Han}, \citenamefont {Chen}, \citenamefont
  {Wang}, \citenamefont {Palma}, \citenamefont {Cheng}, \citenamefont {Lin},
  \citenamefont {Du}, \citenamefont {Ma}, \citenamefont
  {Fern{\'a}ndez-Rossier}, \citenamefont {Feng},\ and\ \citenamefont
  {Gao}}]{fu2024building}%
  \BibitemOpen
  \bibfield  {author} {\bibinfo {author} {\bibfnamefont {X.}~\bibnamefont
  {Fu}}, \bibinfo {author} {\bibfnamefont {L.}~\bibnamefont {Huang}}, \bibinfo
  {author} {\bibfnamefont {K.}~\bibnamefont {Liu}}, \bibinfo {author}
  {\bibfnamefont {J.~C.}\ \bibnamefont {Henriques}}, \bibinfo {author}
  {\bibfnamefont {Y.}~\bibnamefont {Gao}}, \bibinfo {author} {\bibfnamefont
  {X.}~\bibnamefont {Han}}, \bibinfo {author} {\bibfnamefont {H.}~\bibnamefont
  {Chen}}, \bibinfo {author} {\bibfnamefont {Y.}~\bibnamefont {Wang}}, \bibinfo
  {author} {\bibfnamefont {C.-A.}\ \bibnamefont {Palma}}, \bibinfo {author}
  {\bibfnamefont {Z.}~\bibnamefont {Cheng}}, \bibinfo {author} {\bibfnamefont
  {X.}~\bibnamefont {Lin}}, \bibinfo {author} {\bibfnamefont {S.}~\bibnamefont
  {Du}}, \bibinfo {author} {\bibfnamefont {J.}~\bibnamefont {Ma}}, \bibinfo
  {author} {\bibfnamefont {J.}~\bibnamefont {Fern{\'a}ndez-Rossier}}, \bibinfo
  {author} {\bibfnamefont {X.}~\bibnamefont {Feng}},\ and\ \bibinfo {author}
  {\bibfnamefont {H.-J.}\ \bibnamefont {Gao}},\ }\href@noop {} {\bibfield
  {journal} {\bibinfo  {journal} {arXiv preprint arXiv:2407.20511}\ } (\bibinfo
  {year} {2024})}\BibitemShut {NoStop}%
\bibitem [{\citenamefont {Zhao}\ \emph
  {et~al.}(2024{\natexlab{b}})\citenamefont {Zhao}, \citenamefont {Yang},
  \citenamefont {Henriques}, \citenamefont {Ferri-Cort{\'e}s}, \citenamefont
  {Catarina}, \citenamefont {Pignedoli}, \citenamefont {Ma}, \citenamefont
  {Feng}, \citenamefont {Ruffieux}, \citenamefont {Fern{\'a}ndez-Rossier},\
  and\ \citenamefont {Fasel}}]{zhao2024gapless}%
  \BibitemOpen
  \bibfield  {author} {\bibinfo {author} {\bibfnamefont {C.}~\bibnamefont
  {Zhao}}, \bibinfo {author} {\bibfnamefont {L.}~\bibnamefont {Yang}}, \bibinfo
  {author} {\bibfnamefont {J.~C.}\ \bibnamefont {Henriques}}, \bibinfo {author}
  {\bibfnamefont {M.}~\bibnamefont {Ferri-Cort{\'e}s}}, \bibinfo {author}
  {\bibfnamefont {G.}~\bibnamefont {Catarina}}, \bibinfo {author}
  {\bibfnamefont {C.~A.}\ \bibnamefont {Pignedoli}}, \bibinfo {author}
  {\bibfnamefont {J.}~\bibnamefont {Ma}}, \bibinfo {author} {\bibfnamefont
  {X.}~\bibnamefont {Feng}}, \bibinfo {author} {\bibfnamefont {P.}~\bibnamefont
  {Ruffieux}}, \bibinfo {author} {\bibfnamefont {J.}~\bibnamefont
  {Fern{\'a}ndez-Rossier}},\ and\ \bibinfo {author} {\bibfnamefont
  {R.}~\bibnamefont {Fasel}},\ }\href@noop {} {\bibfield  {journal} {\bibinfo
  {journal} {arXiv preprint arXiv:2408.10045}\ } (\bibinfo {year}
  {2024}{\natexlab{b}})}\BibitemShut {NoStop}%
\bibitem [{\citenamefont {Yuan}\ \emph {et~al.}(2025)\citenamefont {Yuan},
  \citenamefont {Zhang}, \citenamefont {Jiang}, \citenamefont {Qian},
  \citenamefont {Wang}, \citenamefont {Liu}, \citenamefont {Liu}, \citenamefont
  {Liu}, \citenamefont {Guan}, \citenamefont {Li}, \citenamefont {Zheng},
  \citenamefont {Liu}, \citenamefont {Jia}, \citenamefont {Qin}, \citenamefont
  {Liu}, \citenamefont {Li},\ and\ \citenamefont {Wang}}]{yuan2025fractional}%
  \BibitemOpen
  \bibfield  {author} {\bibinfo {author} {\bibfnamefont {Z.}~\bibnamefont
  {Yuan}}, \bibinfo {author} {\bibfnamefont {X.-Y.}\ \bibnamefont {Zhang}},
  \bibinfo {author} {\bibfnamefont {Y.}~\bibnamefont {Jiang}}, \bibinfo
  {author} {\bibfnamefont {X.}~\bibnamefont {Qian}}, \bibinfo {author}
  {\bibfnamefont {Y.}~\bibnamefont {Wang}}, \bibinfo {author} {\bibfnamefont
  {Y.}~\bibnamefont {Liu}}, \bibinfo {author} {\bibfnamefont {L.}~\bibnamefont
  {Liu}}, \bibinfo {author} {\bibfnamefont {X.}~\bibnamefont {Liu}}, \bibinfo
  {author} {\bibfnamefont {D.}~\bibnamefont {Guan}}, \bibinfo {author}
  {\bibfnamefont {Y.}~\bibnamefont {Li}}, \bibinfo {author} {\bibfnamefont
  {H.}~\bibnamefont {Zheng}}, \bibinfo {author} {\bibfnamefont
  {C.}~\bibnamefont {Liu}}, \bibinfo {author} {\bibfnamefont {L.}~\bibnamefont
  {Jia}, \bibfnamefont {Jinfeng}}, \bibinfo {author} {\bibfnamefont
  {M.}~\bibnamefont {Qin}}, \bibinfo {author} {\bibfnamefont {Q.}~\bibnamefont
  {Liu}, \bibfnamefont {Pei-Nian}}, \bibinfo {author} {\bibfnamefont {D.-Y.}\
  \bibnamefont {Li}},\ and\ \bibinfo {author} {\bibfnamefont {S.}~\bibnamefont
  {Wang}},\ }\href@noop {} {\bibfield  {journal} {\bibinfo  {journal} {Journal
  of the American Chemical Society}\ } (\bibinfo {year} {2025})}\BibitemShut
  {NoStop}%
\bibitem [{\citenamefont {Su}\ \emph {et~al.}(2024)\citenamefont {Su},
  \citenamefont {Ding}, \citenamefont {Hong}, \citenamefont {Ke}, \citenamefont
  {Yan}, \citenamefont {Li}, \citenamefont {Jiang},\ and\ \citenamefont
  {Yu}}]{su2024fabrication}%
  \BibitemOpen
  \bibfield  {author} {\bibinfo {author} {\bibfnamefont {X.}~\bibnamefont
  {Su}}, \bibinfo {author} {\bibfnamefont {Z.}~\bibnamefont {Ding}}, \bibinfo
  {author} {\bibfnamefont {Y.}~\bibnamefont {Hong}}, \bibinfo {author}
  {\bibfnamefont {N.}~\bibnamefont {Ke}}, \bibinfo {author} {\bibfnamefont
  {K.}~\bibnamefont {Yan}}, \bibinfo {author} {\bibfnamefont {C.}~\bibnamefont
  {Li}}, \bibinfo {author} {\bibfnamefont {Y.}~\bibnamefont {Jiang}},\ and\
  \bibinfo {author} {\bibfnamefont {P.}~\bibnamefont {Yu}},\ }\href@noop {}
  {\bibfield  {journal} {\bibinfo  {journal} {arXiv preprint arXiv:2408.08801}\
  } (\bibinfo {year} {2024})}\BibitemShut {NoStop}%
\bibitem [{\citenamefont {Lovesey}(1984)}]{Lovesey1984}%
  \BibitemOpen
  \bibfield  {author} {\bibinfo {author} {\bibfnamefont {S.~W.}\ \bibnamefont
  {Lovesey}},\ }\href@noop {} {\emph {\bibinfo {title} {Theory of Neutron
  Scattering from Condensed Matter}}},\ \bibinfo {series} {International Series
  of Monographs on Physics}\ No.~\bibinfo {number} {72}\ (\bibinfo  {publisher}
  {Clarendon Press},\ \bibinfo {year} {1984})\BibitemShut {NoStop}%
\bibitem [{\citenamefont {Rossat-Mignod}\ \emph {et~al.}(1991)\citenamefont
  {Rossat-Mignod}, \citenamefont {Regnault}, \citenamefont {Vettier},
  \citenamefont {Bourges}, \citenamefont {Burlet}, \citenamefont {Bossy},
  \citenamefont {Henry},\ and\ \citenamefont {Lapertot}}]{rossat1991neutron}%
  \BibitemOpen
  \bibfield  {author} {\bibinfo {author} {\bibfnamefont {J.}~\bibnamefont
  {Rossat-Mignod}}, \bibinfo {author} {\bibfnamefont {L.}~\bibnamefont
  {Regnault}}, \bibinfo {author} {\bibfnamefont {C.}~\bibnamefont {Vettier}},
  \bibinfo {author} {\bibfnamefont {P.}~\bibnamefont {Bourges}}, \bibinfo
  {author} {\bibfnamefont {P.}~\bibnamefont {Burlet}}, \bibinfo {author}
  {\bibfnamefont {J.}~\bibnamefont {Bossy}}, \bibinfo {author} {\bibfnamefont
  {J.}~\bibnamefont {Henry}},\ and\ \bibinfo {author} {\bibfnamefont
  {G.}~\bibnamefont {Lapertot}},\ }\href@noop {} {\bibfield  {journal}
  {\bibinfo  {journal} {Physica C: Superconductivity}\ }\textbf {\bibinfo
  {volume} {185}},\ \bibinfo {pages} {86} (\bibinfo {year} {1991})}\BibitemShut
  {NoStop}%
\bibitem [{\citenamefont {Benedek}\ and\ \citenamefont
  {Fritsch}(1966)}]{Benedek1966}%
  \BibitemOpen
  \bibfield  {author} {\bibinfo {author} {\bibfnamefont {G.~B.}\ \bibnamefont
  {Benedek}}\ and\ \bibinfo {author} {\bibfnamefont {K.}~\bibnamefont
  {Fritsch}},\ }\href {https://doi.org/10.1103/PhysRev.149.647} {\bibfield
  {journal} {\bibinfo  {journal} {Phys. Rev.}\ }\textbf {\bibinfo {volume}
  {149}},\ \bibinfo {pages} {647} (\bibinfo {year} {1966})}\BibitemShut
  {NoStop}%
\bibitem [{\citenamefont {Dil}(1982)}]{dil1982brillouin}%
  \BibitemOpen
  \bibfield  {author} {\bibinfo {author} {\bibfnamefont {J.}~\bibnamefont
  {Dil}},\ }\href@noop {} {\bibfield  {journal} {\bibinfo  {journal} {Reports
  on Progress in Physics}\ }\textbf {\bibinfo {volume} {45}},\ \bibinfo {pages}
  {285} (\bibinfo {year} {1982})}\BibitemShut {NoStop}%
\bibitem [{\citenamefont {Damascelli}\ \emph {et~al.}(2003)\citenamefont
  {Damascelli}, \citenamefont {Hussain},\ and\ \citenamefont
  {Shen}}]{Damascelli2003}%
  \BibitemOpen
  \bibfield  {author} {\bibinfo {author} {\bibfnamefont {A.}~\bibnamefont
  {Damascelli}}, \bibinfo {author} {\bibfnamefont {Z.}~\bibnamefont
  {Hussain}},\ and\ \bibinfo {author} {\bibfnamefont {Z.-X.}\ \bibnamefont
  {Shen}},\ }\href {https://doi.org/10.1103/RevModPhys.75.473} {\bibfield
  {journal} {\bibinfo  {journal} {Rev. Mod. Phys.}\ }\textbf {\bibinfo {volume}
  {75}},\ \bibinfo {pages} {473} (\bibinfo {year} {2003})}\BibitemShut
  {NoStop}%
\bibitem [{\citenamefont {Sobota}\ \emph {et~al.}(2021)\citenamefont {Sobota},
  \citenamefont {He},\ and\ \citenamefont {Shen}}]{Sobota2021}%
  \BibitemOpen
  \bibfield  {author} {\bibinfo {author} {\bibfnamefont {J.~A.}\ \bibnamefont
  {Sobota}}, \bibinfo {author} {\bibfnamefont {Y.}~\bibnamefont {He}},\ and\
  \bibinfo {author} {\bibfnamefont {Z.-X.}\ \bibnamefont {Shen}},\ }\href
  {https://doi.org/10.1103/RevModPhys.93.025006} {\bibfield  {journal}
  {\bibinfo  {journal} {Rev. Mod. Phys.}\ }\textbf {\bibinfo {volume} {93}},\
  \bibinfo {pages} {025006} (\bibinfo {year} {2021})}\BibitemShut {NoStop}%
\bibitem [{\citenamefont {Petersen}\ \emph {et~al.}(1998)\citenamefont
  {Petersen}, \citenamefont {Sprunger}, \citenamefont {Hofmann}, \citenamefont
  {L\ae{}gsgaard}, \citenamefont {Briner}, \citenamefont {Doering},
  \citenamefont {Rust}, \citenamefont {Bradshaw}, \citenamefont {Besenbacher},\
  and\ \citenamefont {Plummer}}]{Petersen1998}%
  \BibitemOpen
  \bibfield  {author} {\bibinfo {author} {\bibfnamefont {L.}~\bibnamefont
  {Petersen}}, \bibinfo {author} {\bibfnamefont {P.~T.}\ \bibnamefont
  {Sprunger}}, \bibinfo {author} {\bibfnamefont {P.}~\bibnamefont {Hofmann}},
  \bibinfo {author} {\bibfnamefont {E.}~\bibnamefont {L\ae{}gsgaard}}, \bibinfo
  {author} {\bibfnamefont {B.~G.}\ \bibnamefont {Briner}}, \bibinfo {author}
  {\bibfnamefont {M.}~\bibnamefont {Doering}}, \bibinfo {author} {\bibfnamefont
  {H.-P.}\ \bibnamefont {Rust}}, \bibinfo {author} {\bibfnamefont {A.~M.}\
  \bibnamefont {Bradshaw}}, \bibinfo {author} {\bibfnamefont {F.}~\bibnamefont
  {Besenbacher}},\ and\ \bibinfo {author} {\bibfnamefont {E.~W.}\ \bibnamefont
  {Plummer}},\ }\href {https://doi.org/10.1103/PhysRevB.57.R6858} {\bibfield
  {journal} {\bibinfo  {journal} {Phys. Rev. B}\ }\textbf {\bibinfo {volume}
  {57}},\ \bibinfo {pages} {R6858} (\bibinfo {year} {1998})}\BibitemShut
  {NoStop}%
\bibitem [{\citenamefont {Pascual}\ \emph {et~al.}(2004)\citenamefont
  {Pascual}, \citenamefont {Bihlmayer}, \citenamefont {Koroteev}, \citenamefont
  {Rust}, \citenamefont {Ceballos}, \citenamefont {Hansmann}, \citenamefont
  {Horn}, \citenamefont {Chulkov}, \citenamefont {Bl\"ugel}, \citenamefont
  {Echenique},\ and\ \citenamefont {Hofmann}}]{Pascual2004}%
  \BibitemOpen
  \bibfield  {author} {\bibinfo {author} {\bibfnamefont {J.~I.}\ \bibnamefont
  {Pascual}}, \bibinfo {author} {\bibfnamefont {G.}~\bibnamefont {Bihlmayer}},
  \bibinfo {author} {\bibfnamefont {Y.~M.}\ \bibnamefont {Koroteev}}, \bibinfo
  {author} {\bibfnamefont {H.-P.}\ \bibnamefont {Rust}}, \bibinfo {author}
  {\bibfnamefont {G.}~\bibnamefont {Ceballos}}, \bibinfo {author}
  {\bibfnamefont {M.}~\bibnamefont {Hansmann}}, \bibinfo {author}
  {\bibfnamefont {K.}~\bibnamefont {Horn}}, \bibinfo {author} {\bibfnamefont
  {E.~V.}\ \bibnamefont {Chulkov}}, \bibinfo {author} {\bibfnamefont
  {S.}~\bibnamefont {Bl\"ugel}}, \bibinfo {author} {\bibfnamefont {P.~M.}\
  \bibnamefont {Echenique}},\ and\ \bibinfo {author} {\bibfnamefont
  {P.}~\bibnamefont {Hofmann}},\ }\href
  {https://doi.org/10.1103/PhysRevLett.93.196802} {\bibfield  {journal}
  {\bibinfo  {journal} {Phys. Rev. Lett.}\ }\textbf {\bibinfo {volume} {93}},\
  \bibinfo {pages} {196802} (\bibinfo {year} {2004})}\BibitemShut {NoStop}%
\bibitem [{\citenamefont {S\"ode}\ \emph {et~al.}(2015)\citenamefont {S\"ode},
  \citenamefont {Talirz}, \citenamefont {Gr\"oning}, \citenamefont {Pignedoli},
  \citenamefont {Berger}, \citenamefont {Feng}, \citenamefont {M\"ullen},
  \citenamefont {Fasel},\ and\ \citenamefont {Ruffieux}}]{Sode2015}%
  \BibitemOpen
  \bibfield  {author} {\bibinfo {author} {\bibfnamefont {H.}~\bibnamefont
  {S\"ode}}, \bibinfo {author} {\bibfnamefont {L.}~\bibnamefont {Talirz}},
  \bibinfo {author} {\bibfnamefont {O.}~\bibnamefont {Gr\"oning}}, \bibinfo
  {author} {\bibfnamefont {C.~A.}\ \bibnamefont {Pignedoli}}, \bibinfo {author}
  {\bibfnamefont {R.}~\bibnamefont {Berger}}, \bibinfo {author} {\bibfnamefont
  {X.}~\bibnamefont {Feng}}, \bibinfo {author} {\bibfnamefont {K.}~\bibnamefont
  {M\"ullen}}, \bibinfo {author} {\bibfnamefont {R.}~\bibnamefont {Fasel}},\
  and\ \bibinfo {author} {\bibfnamefont {P.}~\bibnamefont {Ruffieux}},\ }\href
  {https://doi.org/10.1103/PhysRevB.91.045429} {\bibfield  {journal} {\bibinfo
  {journal} {Phys. Rev. B}\ }\textbf {\bibinfo {volume} {91}},\ \bibinfo
  {pages} {045429} (\bibinfo {year} {2015})}\BibitemShut {NoStop}%
\bibitem [{\citenamefont {Schouteden}\ \emph {et~al.}(2009)\citenamefont
  {Schouteden}, \citenamefont {Lievens},\ and\ \citenamefont
  {Van~Haesendonck}}]{Schouteden2009}%
  \BibitemOpen
  \bibfield  {author} {\bibinfo {author} {\bibfnamefont {K.}~\bibnamefont
  {Schouteden}}, \bibinfo {author} {\bibfnamefont {P.}~\bibnamefont
  {Lievens}},\ and\ \bibinfo {author} {\bibfnamefont {C.}~\bibnamefont
  {Van~Haesendonck}},\ }\href {https://doi.org/10.1103/PhysRevB.79.195409}
  {\bibfield  {journal} {\bibinfo  {journal} {Phys. Rev. B}\ }\textbf {\bibinfo
  {volume} {79}},\ \bibinfo {pages} {195409} (\bibinfo {year}
  {2009})}\BibitemShut {NoStop}%
\bibitem [{\citenamefont {Appelbaum}(1967)}]{Appelbaum1967Exchange}%
  \BibitemOpen
  \bibfield  {author} {\bibinfo {author} {\bibfnamefont {J.~A.}\ \bibnamefont
  {Appelbaum}},\ }\href@noop {} {\bibfield  {journal} {\bibinfo  {journal}
  {Phys. Rev.}\ }\textbf {\bibinfo {volume} {154}},\ \bibinfo {pages} {633}
  (\bibinfo {year} {1967})}\BibitemShut {NoStop}%
\bibitem [{\citenamefont {Fern\'andez-Rossier}(2009)}]{Rossier2009ITS}%
  \BibitemOpen
  \bibfield  {author} {\bibinfo {author} {\bibfnamefont {J.}~\bibnamefont
  {Fern\'andez-Rossier}},\ }\href@noop {} {\bibfield  {journal} {\bibinfo
  {journal} {Physical Review Letters}\ }\textbf {\bibinfo {volume} {102}},\
  \bibinfo {pages} {256802} (\bibinfo {year} {2009})}\BibitemShut {NoStop}%
\bibitem [{\citenamefont {Ternes}(2015)}]{ternes2015spin}%
  \BibitemOpen
  \bibfield  {author} {\bibinfo {author} {\bibfnamefont {M.}~\bibnamefont
  {Ternes}},\ }\href@noop {} {\bibfield  {journal} {\bibinfo  {journal} {New
  Journal of Physics}\ }\textbf {\bibinfo {volume} {17}},\ \bibinfo {pages}
  {063016} (\bibinfo {year} {2015})}\BibitemShut {NoStop}%
\bibitem [{SM()}]{SM}%
  \BibitemOpen
  \href@noop {} {}\bibinfo {note} {See Supplemental Material at [URL or DOI, if
  available] for additional details, figures, and derivations.}\BibitemShut
  {Stop}%
\bibitem [{\citenamefont {Collins}\ \emph {et~al.}(2006)\citenamefont
  {Collins}, \citenamefont {Hamer},\ and\ \citenamefont
  {Weihong}}]{Collins2006}%
  \BibitemOpen
  \bibfield  {author} {\bibinfo {author} {\bibfnamefont {A.}~\bibnamefont
  {Collins}}, \bibinfo {author} {\bibfnamefont {C.~J.}\ \bibnamefont {Hamer}},\
  and\ \bibinfo {author} {\bibfnamefont {Z.}~\bibnamefont {Weihong}},\ }\href
  {https://doi.org/10.1103/PhysRevB.74.144414} {\bibfield  {journal} {\bibinfo
  {journal} {Phys. Rev. B}\ }\textbf {\bibinfo {volume} {74}},\ \bibinfo
  {pages} {144414} (\bibinfo {year} {2006})}\BibitemShut {NoStop}%
\bibitem [{\citenamefont {Kittel}(1991)}]{kittel1991quantum}%
  \BibitemOpen
  \bibfield  {author} {\bibinfo {author} {\bibfnamefont {C.}~\bibnamefont
  {Kittel}},\ }\href
  {https://www.wiley.com/en-us/Quantum+Theory+of+Solids%2C+2nd+Revised+Edition-p-9780471624127}
  {\emph {\bibinfo {title} {Quantum Theory of Solids}}},\ \bibinfo {edition}
  {2nd}\ ed.\ (\bibinfo  {publisher} {Wiley},\ \bibinfo {address} {New York},\
  \bibinfo {year} {1991})\BibitemShut {NoStop}%
\bibitem [{\citenamefont {Krane}\ \emph {et~al.}(2024)\citenamefont {Krane},
  \citenamefont {Turco}, \citenamefont {Bernhardt}, \citenamefont
  {Jur{\'\i}{\v{c}}ek}, \citenamefont {Fasel},\ and\ \citenamefont
  {Ruffieux}}]{krane2024vibrational}%
  \BibitemOpen
  \bibfield  {author} {\bibinfo {author} {\bibfnamefont {N.}~\bibnamefont
  {Krane}}, \bibinfo {author} {\bibfnamefont {E.}~\bibnamefont {Turco}},
  \bibinfo {author} {\bibfnamefont {A.}~\bibnamefont {Bernhardt}}, \bibinfo
  {author} {\bibfnamefont {M.}~\bibnamefont {Jur{\'\i}{\v{c}}ek}}, \bibinfo
  {author} {\bibfnamefont {R.}~\bibnamefont {Fasel}},\ and\ \bibinfo {author}
  {\bibfnamefont {P.}~\bibnamefont {Ruffieux}},\ }\href@noop {} {\bibfield
  {journal} {\bibinfo  {journal} {arXiv preprint arXiv:2411.19670}\ } (\bibinfo
  {year} {2024})}\BibitemShut {NoStop}%
\end{thebibliography}

%

\end{document}